\documentclass[12pt]{article}%
\usepackage{amsfonts}
\usepackage{amsmath}
\usepackage{amssymb}
\usepackage{graphicx}%
\providecommand{\U}[1]{\protect\rule{.1in}{.1in}}

\begin{document}

\title{Generalized forms, metrics and Ricci flat four-metrics}
\author{D C Robinson\\Mathematics Department\\King's College London\\Strand\\London WC2R 2LS\\United Kingdom\\email: david.c.robinson@kcl.ac.uk}
\maketitle

\begin{abstract}
Generalized differential forms are used in discussions of metric geometries
and Einstein's vacuum field equations. \ Cartan's structure equations are
generalized and applied. \ In particular flat generalized connections are
associated with any metric. \ Einstein's vacuum field equations and its
Lorentzian four-metric solutions are considered from a novel point of view.
\ It is shown how particular generalized flat geometries can encode such Ricci
flat metrics.

\end{abstract}

\section{Introduction}

When a differential form of negative degree, \cite{sp1}, is adjoined to
ordinary differential forms an extension of the standard exterior algebra and
calculus can be formulated. \ Such a development was initiated in \cite{rob1}
when a single form of negative degree, minus one, was so adjoined. \ This was
then extended by adjoining $N$ linearly independent negative degree forms.
\ The algebra and calculus of type $N\geqq1$ differential forms is described
in \cite{rob3} and its references. \ Generalized differential forms have been
used to study classical field theories including Yang-Mills, general
relativity, generalized Chern-Simons action principles and higher Yang-Mills,
for example as in \cite{tung1}, \cite{tung2}, \cite{rob4}, \cite{song}. \ The
main aim of this paper is to use generalized forms to investigate ordinary
geometries. \ First, it will be shown how ordinary metric geometries can be
represented by generalized forms and flat type $N=1$ generalized connections.
\ Second it will be demonstrated how\ the condition that a Lorentzian
four-metric satisfies Einstein's vacuum field equations with zero cosmological
constant, in other words be Ricci flat, can be encoded using generalized forms
and a type $N=2$ flat generalized connection\footnote{This paper deals with
local formulations. \ Generalized forms are denoted with bold Latin letters
and ordinary forms with Greek letters. \ Occasionally the degree of a form is
denoted above it. \ Lower case Latin\ indices range and sum over 1 to the
dimension of the manifold being considered. Upper case spinor indices range
and sum over 0-1.}.

More specifically, by using type $N=1$ generalized forms it will be shown how
a co-frame $\{\theta^{a}\}$ for a metric of signature $(p,q),$ $ds^{2}%
=\eta_{ab}\theta^{a}\otimes\theta^{b}$ where $\eta_{ab}$ is constant, can
always be expressed as $\theta^{a}=(\mathbf{L}^{-1})_{b}^{a}d\mathbf{x}^{b}$
where the generalized zero-forms $\mathbf{L}_{b}^{a}$ satisfy $\mathbf{L}%
_{c}^{a}\mathbf{L}_{d}^{b}\eta_{ab}=\eta_{cd}$. \ The degree zero generalized
forms $\{\mathbf{x}^{b}\}$ can be thought of as generalized form (gf-)
coordinates \cite{rob1}. \ Hence any metric can be formally expressed in terms
of a flat generalized line element\textit{, }$ds^{2}=\eta_{ab}d\mathbf{x}%
^{a}\otimes d\mathbf{x}^{b}$. \ This is in contrast to the standard local
result for ordinary forms that a metric admits co-frames which can be
expressed, in appropriate local coordinates $\{x^{a}\},$ as $\theta
^{a}=(L^{-1})_{b}^{a}dx^{b},$ where $L_{b}^{a}\in SO(p,q),$ if and only if the
metric is flat. \ It will also be shown that an ordinary Lorentzian 4-metric,
$ds^{2}=\eta_{ab}\theta^{a}\otimes\theta^{b},$ is Ricci flat if and only if
$\theta^{a}=(\mathbf{L}^{-1})_{b}^{a}d\mathbf{x}^{b}$ where $\mathbf{L}%
_{b}^{a}$ and $\mathbf{x}^{b}$ are certain type $N=2$ zero-forms and
$\mathbf{L}_{c}^{a}\mathbf{L}_{d}^{b}\eta_{ab}=\eta_{cd}$.\ \ These results
extend a line of previous work in which both the Yang-Mills and Einstein's
vacuum gravitational field equations were shown to be representable by flat
generalized connections \cite{rob6}, \cite{rob7}. \ \ Although four-metrics of
Lorentzian signature\ are of particular interest in this paper many of the
results presented here can be similarly applied to metric geometries of other
signatures and in other dimensions.

The paper is structured as follows. \ In the second section Cartan's structure
equations for ordinary metric geometries are reviewed in order to exhibit the
notation that is being used. \ These equations are then extended to apply to
generalized forms. \ Applications of the formalism developed here are
considered in the following sections. \ The third section deals with Cartan's
equations and type $N=1$ forms. \ In particular it is shown that any ordinary
metric can be expressed in a generalized flat form. \ The fourth section
reviews the two-component spinor formalism for ordinary and generalized forms
which it is useful, although not essential, to use in the following sections
\cite{pen1}, \cite{rob9}. \ It deals with the particular case of Lorentzian
four-metrics, the corresponding Cartan structure equations and Einstein's
vacuum field equations. \ The formalism of section four is used in the fifth
section where the ordinary Cartan's first structure equations are extended to
equations which are satisfied if and only if the corresponding ordinary
Lorentzian four-metric is Ricci flat. \ This is done by using type $N=2$ forms
and a certain flat generalized connection. \ It is shown how any Ricci flat
ordinary Lorentzian four-metric, $g$, determines and is determined by
generalized form (gf-) coordinates $\{\mathbf{x}^{a}\}$ and co-frames,
$\theta^{a}=(\mathbf{L}^{-1})_{b}^{a}d\mathbf{x}^{b}$ and hence that $g$ can
be formally expressed by the line element $ds^{2}=\eta_{ab}d\mathbf{x}%
^{a}\otimes d\mathbf{x}^{b}.$ \ The freedom in the choice of gf-coordinates
$\{\mathbf{x}^{a}\}$ is expressed in terms of a generalization of the
Poincar\'{e} group. \ The final section contains a brief discussion of the
results in this paper.

\section{\textbf{Generalized Cartan structure equations for type }%
$N=2$\textbf{ generalized forms}}

Recall first the Cartan structure equations for an ordinary metric $g$ on an
$n$ dimensional manifold $M$, with line element%
\begin{equation}
ds^{2}=\eta_{ab}\theta^{a}\otimes\theta^{b},
\end{equation}
where $\{\theta^{a}\}$is a basis of ordinary one-forms with respect to which
the metric, $g$, has constant components $\eta_{ab}$. \ When the metric
$\eta_{ab}$ has signature $(p,q)$ the corresponding ordinary structure group
is $G=SO(p,q)$, with Lie algebra $\mathfrak{g=so(p,q)}$.

The first set of Cartan structure equations is
\begin{equation}
D\theta^{a}\equiv d\theta^{a}+\omega_{b}^{a}\theta^{b}=0
\end{equation}
and the second set is%
\begin{equation}
\Omega_{b}^{a}=d\omega_{b}^{a}+\omega_{c}^{a}\omega_{b}^{c}.
\end{equation}
Here\ $\omega_{b}^{a}$ is the torsion free and metric Levi-Civita connection,
and its curvature two-form is $\Omega_{b}^{a}\footnote{The focus in this paper
will be on Cartan's equations with torsion free metric connections and their
generalizations. \ It is a straightforward matter to extend the investigations
to other connections.}$. \ The symbol $D$ denotes the Levi-Civita covariant
exterior derivative. \ The first and second Bianchi identities are%
\begin{equation}
\Omega_{b}^{a}\theta^{b}=0;\text{ }D\Omega_{b}^{a}=0.
\end{equation}

Under a $SO(p,q)$ gauge transformation $L_{b}^{a}$ satisfying%
\begin{equation}
L_{c}^{a}L_{d}^{b}\eta_{ab}=\eta_{cd}%
\end{equation}%
\begin{align}
\theta^{a}  &  \rightarrow\theta_{1}^{a}=(L^{-1})_{b}^{a}\theta^{b},\\
\omega_{b}^{a}  &  \rightarrow\omega_{1b}^{a}=(L^{-1})_{c}^{a}dL_{b}%
^{c}+(L^{-1})_{c}^{a}\omega_{d}^{c}L_{b}^{d},\nonumber\\
\Omega_{b}^{a}  &  \rightarrow\Omega_{1b}^{a}=(L^{-1})_{c}^{a}\Omega_{d}%
^{c}L_{b}^{d}.\nonumber
\end{align}
The curvature vanishes and the connection is flat if and only if there exist
local coordinates $\{x^{a}\}$ such that any co-frame satisfying Cartan's
equations is locally given by%
\begin{equation}
\theta^{a}=(L^{-1})_{b}^{a}dx^{b},
\end{equation}
for some $L_{b}^{a}\in SO(p,q),$ and then $ds^{2}=\eta_{ab}dx^{a}\otimes
dx^{b}.$

Following previous papers the type $N=2$ generalized forms to be used in this
paper will be constructed by adjoining to ordinary forms a complex conjugate
pair of minus one-forms $\mathbf{m}$ and $\overline{\mathbf{m}}$, where
$\mathbf{m}\overline{\mathbf{m}}$ is non-zero. \ These minus one forms obey
the usual rules of exterior algebra and calculus and hence so do all the
generalized forms\footnote{Generalized forms, their algebra and calculus,
together with generalized vector fields, admit representations in terms of
functions and vector fields on the Whitney sum of the reverse parity tangent
bundle, $\Pi TM$, of $M$ and a trivial reverse parity bundle over $M$. \ \ For
type $N=2$ forms, which are the type mainly considered in appplications in
this paper, the latter, has fibre $R^{0\mid2}$, that is $R^{2}$ with parity
reversed. \ When complex forms are used the fibre is $C$ with parity reversed,
that is $C^{0\mid1}$. \ Other type $N$ forms and generalized vector fields can
be represented by similar generalizations of the known representation of
ordinary forms and vector fields \cite{rob8}, \cite{chat}.}. \ Their exterior
derivatives will be chosen to be
\begin{equation}
d\mathbf{m}\mathbf{=d}\overline{\mathbf{m}}=1,\text{ }d^{2}=0.
\end{equation}
Any type $N=2$ generalized p-form, $\overset{p}{\mathbf{p}}$ can be expressed
in terms of an expansion in these and ordinary forms as%
\begin{equation}
\overset{p}{\mathbf{p}}=\overset{p}{\pi}+\overset{p+1}{\pi_{1}}\mathbf{m}%
+\overset{p+1}{\pi_{2}}\overline{\mathbf{m}}+\overset{p+2}{\pi}\mathbf{m}%
\overline{\mathbf{m}}.
\end{equation}
When $\overset{p}{\mathbf{p}}$ is real, $\overset{p}{\pi}$ is real,
$\overset{p+1}{\pi_{2}}$ is the complex conjugate of $\overset{p+1}{\pi_{1}}$
and $\overset{p+2}{\pi}$ is purely imaginary.

If $\overset{q}{\mathbf{q}}=\overset{q}{\upsilon}+\overset{q+1}{\upsilon_{1}%
}\mathbf{m}+\overset{q+1}{\upsilon_{2}}\overline{\mathbf{m}}%
+\overset{q+2}{\upsilon}\mathbf{m}\overline{\mathbf{m}}$,%
\begin{align*}
\overset{p}{\mathbf{p}}\overset{q}{\mathbf{q}}  &  =\overset{p}{\pi
}\overset{q}{\upsilon}+[\overset{p}{\pi}\overset{q+1}{\upsilon}_{1}%
+(-1)^{q}\overset{p+1}{\pi}_{1}\overset{q}{\upsilon}]\mathbf{m+}%
[\overset{p}{\pi}\overset{q+1}{\upsilon}_{2}+(-1)^{q}\overset{p+1}{\pi}%
_{2}\overset{q}{\upsilon}]\overline{\mathbf{m}}\\
&  +[\overset{p}{\pi}\overset{q+2}{\upsilon}+(-1)^{q+1}\overset{p+1}{\pi}%
_{1}\overset{q+1}{\upsilon}_{2}+(-1)^{q}\overset{p+1}{\pi}_{2}%
\overset{q+1}{\upsilon}_{1}+\overset{p+2}{\pi}\overset{q}{\upsilon}%
]\mathbf{m}\overline{\mathbf{m}},\\
d\overset{p}{\mathbf{p}}  &  =d\overset{p}{\pi}+(-1)^{p+1}(\overset{p+1}{\pi
_{1}}+\overset{p+1}{\pi_{2}})+[d\overset{p+1}{\pi_{1}}+(-1)^{p+1}%
\overset{p+2}{\pi}]\mathbf{m}+[\overset{p+1}{d\pi_{2}}+(-1)^{p}%
\overset{p+2}{\pi}]\overline{\mathbf{m}}\\
&  +d\overset{p+2}{\pi}\mathbf{m}\overline{\mathbf{m}},
\end{align*}
$\overset{p}{\mathbf{p}}\overset{q}{\mathbf{q}}=(-1)^{pq}%
\overset{q}{\mathbf{q}}\overset{p}{\mathbf{p}}$, $d(\overset{p}{\mathbf{p}%
}\overset{q}{\mathbf{q}})=(d\overset{p}{\mathbf{p}})\overset{q}{\mathbf{q}%
}+(-1)^{p}\overset{p}{\mathbf{p}}d\overset{q}{\mathbf{q}}$ , and $d^{2}=0.$

Sometimes it is convenient to use the real and imaginary parts of$\mathbf{\ m}%
$ as basis minus one-forms. For this basis of minus one-forms $\{\mathbf{m}%
^{1},\mathbf{m}^{2}\}$
\begin{align}
\mathbf{m}^{1}  &  =\overline{\mathbf{m}}^{1},\mathbf{m}^{2}=\overline
{\mathbf{m}}^{2}\text{; }\mathbf{m=\mathbf{m}^{1}+i\mathbf{m}^{2},}\\
d\mathbf{m}^{1}  &  =1,d\mathbf{m}^{2}=0.\nonumber
\end{align}

When the ordinary forms are replaced by generalized forms in Eqs.(2) and (3)
the corresponding generalized Cartan equations so obtained are%
\begin{align}
\mathbf{D\mathbf{e}}^{a}  &  =d\mathbf{\mathbf{e}}^{a}+\mathbf{A}_{b}%
^{a}\mathbf{\mathbf{e}}^{b}=0,\\
\mathbf{F}_{b}^{a}  &  =d\mathbf{A}_{b}^{a}+\mathbf{A}_{c}^{a}\mathbf{A}%
_{b}^{c},\nonumber
\end{align}
where $\mathbf{A}_{ab}+\mathbf{A}_{ba}=0$, $\mathbf{A}_{ab}=\eta
_{ac}\mathbf{A}_{b}^{c}$, and $\mathbf{A}_{b}^{a}$ is a generalized connection
with covariant exterior derivative $\mathbf{D}$.

The first and second generalized Bianchi identities are%
\begin{equation}
\mathbf{F}_{b}^{a}\mathbf{\mathbf{e}}^{b}=0;\text{ }\mathbf{DF}_{b}^{a}=0.
\end{equation}

The natural extensions of the ordinary gauge transformations, Eq.(6), are%

\begin{align}
\mathbf{e}^{a}  &  \rightarrow(\mathbf{L}^{-1})_{b}^{a}\mathbf{e}^{b},\\
\mathbf{A}_{b}^{a}  &  \rightarrow(\mathbf{L}^{-1})_{c}^{a}d\mathbf{L}_{b}%
^{c}+(\mathbf{L}^{-1})_{c}^{a}\mathbf{A}_{d}^{c}\mathbf{L}_{b}^{d},\nonumber\\
\mathbf{F}_{b}^{a}  &  \rightarrow(\mathbf{L}^{-1})_{c}^{a}\mathbf{F}_{d}%
^{c}L_{b}^{d}.\nonumber
\end{align}
where the generalized zero-forms $\mathbf{L}_{b}^{a}$ satisfy the $G=$
$SO(p,q)$ type conditions%
\begin{equation}
\mathbf{L}_{b}^{a}\eta_{ac}\mathbf{L}_{d}^{c}=\eta_{bd}.
\end{equation}
The matrix-valued zero forms represented by $\mathbf{L}_{b}^{a}$ constitute,
pointwise, a group (denoted $\mathbf{G}_{(N)}$ for type $N$ forms) under form
-valued matrix multiplication \cite{rob3}. \ More explicitly here,
$\mathbf{G}_{(2)}=\{\mathbf{L}_{b}^{a}\}$ where%
\begin{align}
\mathbf{L}_{b}^{a}  &  =(\delta_{c}^{a}+\mathbf{l}_{c}^{a}\mathbf{m}%
^{2})(\delta_{d}^{c}+_{1}\lambda_{d}^{c}\mathbf{m}^{1})L_{b}^{d},\\
\mathbf{l}_{c}^{a}  &  =_{2}\lambda_{c}^{a}+\lambda_{c}^{a}\mathbf{m}%
^{1},\nonumber
\end{align}
and the matrices $L,$ $_{1}\lambda,$ $_{2}\lambda,$ $\lambda$ respectively
denote a $G$-valued function, two ordinary $\mathfrak{g}$-valued one-forms and
an ordinary $\mathfrak{g}$-valued two-form where $\mathfrak{g}$ is the Lie
algebra of $G$.

The expansion in terms of ordinary forms of the generalized Cartan equations
on an $n$ dimensional manifold when $N=2$ is the following. \ Let%
\begin{align}
\mathbf{\mathbf{e}}^{a}  &  \mathbf{=}\theta^{a}-\tau^{a}\mathbf{m}%
-\overline{\tau}^{a}\overline{\mathbf{m}}+\zeta^{a}\mathbf{m}\overline
{\mathbf{m}};\\
\mathbf{A}_{b}^{a}  &  =\alpha_{b}^{a}-\beta_{b}^{a}\mathbf{m}-\overline
{\beta}_{b}^{a}\overline{\mathbf{m}}+\gamma_{b}^{a}\mathbf{m}\overline
{\mathbf{m}}.\nonumber
\end{align}
where the ordinary one-forms $\theta^{a}$, assumed in this paper to constitute
a co-frame on $M$, and $\alpha_{b}^{a}$ are real, $\tau^{a}$ and $\beta
_{b}^{a}$ are complex, and $\zeta^{a}$ and $\gamma_{b}^{a}$ are pure
imaginary. Then the first generalized Cartan equations given in Eq.(11) are%
\begin{align}
D\theta^{a}  &  =\tau^{a}+\overline{\tau}^{a}\mathbf{,}\\
D\tau^{a}-\zeta^{a}-\beta_{b}^{a}\theta^{b}  &  =0,\nonumber\\
D\zeta^{a}+\beta_{b}^{a}\overline{\tau}^{b}-\overline{\beta}_{b}^{a}\tau
^{b}+\gamma_{b}^{a}\theta^{b}  &  =0,\nonumber
\end{align}
where $D$ denotes the covariant exterior derivative with respect to
$\alpha_{b}^{a}$. \ From these equations it is a straightforward matter to
deduce the equivalent set%
\begin{align}
D\theta^{a}  &  =\tau^{a}+\overline{\tau}^{a}\mathbf{,}\\
\lbrack\digamma_{b}^{a}-(\beta_{b}^{a}+\overline{\beta}_{b}^{a})]\theta^{b}
&  =0,\nonumber\\
D(\tau^{a}-\overline{\tau}^{a})+(\overline{\beta}_{b}^{a}-\beta_{b}^{a}%
)\theta^{b}  &  =2\zeta^{a},\nonumber\\
D\zeta^{a}+\beta_{b}^{a}\overline{\tau}^{b}-\overline{\beta}_{b}^{a}\tau
^{b}+\gamma_{b}^{a}\theta^{b}  &  =0,\nonumber
\end{align}
where \
\begin{equation}
\digamma_{b}^{a}=d\alpha_{b}^{a}+\alpha_{c}^{a}\alpha_{b}^{c},
\end{equation}
and D denotes the covariant exterior derivative with respect to $\alpha
_{b}^{a}$. \ 

The second of Eqs.(11) is given by%
\begin{align}
\mathbf{F}_{b}^{a}  &  =d\mathbf{A}_{b}^{a}+\mathbf{A}_{c}^{a}\mathbf{A}%
_{b}^{c}\\
&  =\digamma_{b}^{a}-(\beta_{b}^{a}+\overline{\beta}_{b}^{a})-(D\beta_{b}%
^{a}-\gamma_{b}^{a})\mathbf{m-(}D\overline{\beta}_{b}^{a}+\gamma_{b}%
^{a})\overline{\mathbf{m}}\nonumber\\
&  +(D\gamma_{b}^{a}+\beta_{c}^{a}\overline{\beta}_{b}^{c}-\overline{\beta
}_{c}^{a}\beta_{b}^{c})\mathbf{m}\overline{\mathbf{m}}.\nonumber
\end{align}
\ The expansions of the generalized Cartan equations, Eqs.(18), are satisfied
if and only if the following hold. \ From the first of Eqs.(18) $\tau
^{a}+\overline{\tau}^{a}$ is the torsion of the ordinary $\mathfrak{so(p,q)}%
$-valued connection given by $\alpha_{b}^{a}$. \ From the second%
\begin{align}
(\beta_{b}^{a}+\overline{\beta}_{b}^{a})  &  =\digamma_{b}^{a}-\mathfrak{f}%
_{b}^{a}\\
\mathfrak{f}_{b}^{a}  &  =\frac{1}{2}\mathfrak{f}_{bcd}^{a}\theta^{c}%
\theta^{d},\nonumber
\end{align}
where\ $\mathfrak{f}_{b}^{a}\theta^{b}=0$, that is the functions
$\mathfrak{f}_{bcd}^{a}$ satisfy the conditions%
\begin{align}
\mathfrak{f}_{abcd}  &  =\mathfrak{f}_{[ab]cd}=\mathfrak{f}_{ab[cd]},\\
\mathfrak{f}_{[bcd]}^{a}  &  =0.\nonumber
\end{align}
The third of Eqs.(18) merely determine the three-forms $\zeta^{a}$.
\ Inserting these results in the fourth equation gives%
\begin{equation}
2\gamma_{b}^{a}\theta^{b}=(\overline{\tau}^{b}-\tau^{b})\mathfrak{f}_{b}%
^{a}+D(\beta_{b}^{a}-\overline{\beta}_{b}^{a})\theta^{b}.
\end{equation}
The expanded curvature two-form is now%
\begin{align}
\mathbf{F}_{b}^{a}  &  =\mathfrak{f}+[\frac{1}{2}D\mathfrak{f}_{b}^{a}%
-\frac{1}{2}D(\beta_{b}^{a}-\overline{\beta}_{b}^{a})+\gamma_{b}%
^{a}]\mathbf{m+}[\frac{1}{2}D\mathfrak{f}_{b}^{a}+\frac{1}{2}D(\beta_{b}%
^{a}-\overline{\beta}_{b}^{a})-\gamma_{b}^{a}]\overline{\mathbf{m}}\\
&  +[D\gamma_{b}^{a}+\frac{1}{2}(\beta_{c}^{a}-\overline{\beta}_{c}^{a}%
)(F_{b}^{c}-f_{b}^{c})-\frac{1}{2}(F_{c}^{a}-f_{c}^{a})(\beta_{b}%
^{c}-\overline{\beta}_{b}^{c})]\mathbf{m}\overline{\mathbf{m}}.\nonumber
\end{align}
In summary: the $N=2$ Cartan structure equations, Eqs.(18)-(20), are satisfied
by%
\begin{align}
\mathbf{\mathbf{e}}^{a}  &  \mathbf{=}\theta^{a}-\tau^{a}\mathbf{m}%
-\overline{\tau}^{a}\overline{\mathbf{m}}+\frac{1}{2}[D(\tau^{a}%
-\overline{\tau}^{a})+(\overline{\beta}_{b}^{a}-\beta_{b}^{a})\theta
^{b}]\mathbf{m}\overline{\mathbf{m}},\\
\mathbf{A}_{b}^{a}  &  =\alpha_{b}^{a}-\beta_{b}^{a}\mathbf{m}-\overline
{\beta}_{b}^{a}\overline{\mathbf{m}}+\gamma_{b}^{a}\mathbf{m}\overline
{\mathbf{m}},\nonumber
\end{align}
where $\tau^{a}+\overline{\tau}^{a}$ is the torsion of the connection given by
$\alpha_{b}^{a}$, $(\beta_{b}^{a}+\overline{\beta}_{b}^{a})=\digamma_{b}%
^{a}-\mathfrak{f}_{b}^{a}$, with $\mathfrak{f}_{b}^{a}$ satisfying the
conditions in Eq.(22), $Im\tau_{b}^{a},$ $Im\beta_{b}^{a}$ and $\gamma_{b}%
^{a}$ being related by Eq.(23), and the curvature two-form being given by Eq.(24).

The curvature two-form given in Eq.(20) vanishes, that is the connection
$\mathbf{A}_{b}^{a}$ is flat, if and only if
\begin{equation}
\digamma_{b}^{a}-(\beta_{b}^{a}+\overline{\beta}_{b}^{a})=0;\text{ }\frac
{1}{2}D(\beta_{b}^{a}-\overline{\beta}_{b}^{a})=\gamma_{b}^{a}.
\end{equation}
\ Then
\begin{equation}
\mathbf{A}_{b}^{a}=\alpha_{b}^{a}-\frac{1}{2}(\digamma_{b}^{a}+\beta_{b}%
^{a}-\overline{\beta}_{b}^{a})\mathbf{m}-\frac{1}{2}(\digamma_{b}^{a}%
-\beta_{b}^{a}+\overline{\beta}_{b}^{a})\overline{\mathbf{m}}+\frac{1}%
{2}D(\beta_{b}^{a}-\overline{\beta}_{b}^{a})\mathbf{m}\overline{\mathbf{m}}.
\end{equation}
By expressing the generalized forms in terms of their expansions in ordinary
forms it can be shown that such a flat connection can be expressed as%
\begin{equation}
\mathbf{A}_{b}^{a}=(\mathbf{L}^{-1})_{c}^{a}d\mathbf{L}_{b}^{c}\mathbf{,}%
\end{equation}
where the matrix of generalized zero-forms belongs to $\mathbf{G}_{(2)}$ as in
Eq.(15) above. \ Such a matrix is not unique since if $_{c}\mathbf{L}_{b}^{e}$
is any closed (and hence exact) element of $\mathbf{G}_{(2)}$ $(\mathbf{L}%
^{-1})_{c}^{a}d\mathbf{L}_{b}^{c}=(\mathbf{L}^{-1})_{d}^{a}(_{c}%
\mathbf{L}^{-1})_{e}^{d}d(_{c}\mathbf{L}_{f}^{e}\mathbf{L}_{b}^{f})$. \ Such
matrices are given by
\begin{equation}
\mathbf{L}_{b}^{a}=\delta_{b}^{a}-\frac{1}{2}(\alpha_{b}^{a}+i\lambda_{2b}%
^{a})\mathbf{m-}\frac{1}{2}(\alpha_{b}^{a}-i\lambda_{2b}^{a})\overline
{\mathbf{m}}+\frac{1}{2}(\beta_{b}^{a}-\overline{\beta}_{b}^{a}-id\lambda
_{2b}^{a}-i\lambda_{2c}^{a}\alpha_{b}^{c})\mathbf{m}\overline{\mathbf{m}},
\end{equation}
where $\lambda_{2b}^{a}$ denotes an arbitrary, ordinary, $\mathfrak{g}$-valued one-form.

When the second generalized Cartan equations are satisfied by a flat
$\mathfrak{so(p,q)}_{2}$-valued connection so that $\mathbf{A}_{b}%
^{a}=(\mathbf{L}^{-1})_{c}^{a}d\mathbf{L}_{b}^{c},$ as in Eq.(28), it then
follows that the first generalized Cartan equations are satisfied by the
generalized one-forms $\mathbf{e}^{a}=(\mathbf{L}^{-1})_{b}^{a}d\mathbf{x}%
^{b}$ where $\mathbf{x}^{b}$ are generalized zero-forms. \ For these solutions
it follows from Eqs.(21) \& (26) that $\mathfrak{f}_{b}^{a}=0$, Eq.(23) is
automatically satisfied with $2Re\tau^{a}$ the torsion of the ordinary
connection $\alpha_{b}^{a}$, $2Re\beta_{b}^{a}$ the curvature two-form of
$\alpha_{b}^{a}$, and Im$\tau^{a}$ and $\operatorname{Im}\beta_{b}^{a}$
arbitrary two-forms. \ The following sections contain applications and more
detailed explorations of these results in some special cases.

\section{Type $N=1$ forms and generalized flatness}

Type $N=1$ forms can be treated as a special case of type $N=2$ forms by
setting $\mathbf{m=}\overline{\mathbf{m}}$. \ Hence a real type $N=1$ form can
be expressed as $\overset{p}{\mathbf{p}}=\overset{p}{\pi}+\overset{p+1}{\pi
}\mathbf{m}^{1}$ where the ordinary forms are real. \ Ordinary forms are type
$N=0$ forms.

The expanded Cartan equations for type $N=1$ forms, \cite{rob1}, can be
extracted from the $N=2$ equations above by setting $\mathbf{m}=\overline
{\mathbf{m}}$, $\tau^{a}=\overline{\tau}^{a}$, $\zeta^{a}=0$, $\beta_{b}%
^{a}=\overline{\beta}_{b}^{a}$, $\gamma_{b}^{a}=0.$ \ Hence, absorbing factors
of 2 in $\tau^{a}$ and $\beta_{b}^{a}$, the type $N=1$ generalized Cartan
equations, $\mathbf{D\mathbf{e}}^{a}=d\mathbf{\mathbf{e}}^{a}+\mathbf{A}%
_{b}^{a}\mathbf{\mathbf{e}}^{b}=0$ and $\mathbf{F}_{b}^{a}=d\mathbf{A}_{b}%
^{a}+\mathbf{A}_{c}^{a}\mathbf{A}_{b,}^{c}$ with%
\begin{align}
\mathbf{\mathbf{e}}^{a}  &  \mathbf{=}\theta^{a}-\tau^{a}\mathbf{m}\\
\mathbf{A}_{b}^{a}  &  =\alpha_{b}^{a}-\beta_{b}^{a}\mathbf{m}\nonumber
\end{align}
expand to become%
\begin{align}
D\theta^{a}  &  =\tau^{a},\\
D\tau^{a}  &  =\beta_{b}^{a}\theta^{b},\nonumber\\
\mathbf{F}_{b}^{a}  &  =\digamma_{b}^{a}-\beta_{b}^{a}-D\beta_{b}%
^{a}\mathbf{m,}\nonumber\\
\digamma_{b}^{a}  &  =d\alpha_{b}^{a}+\alpha_{c}^{a}\alpha_{b}^{c}.\nonumber
\end{align}
When $n>2$, and assuming that $\{\theta^{a}\}$ is an ordinary co-frame, it
follows from Eqs. (31) and their covariant exterior derivatives that, as in
the previous section,%
\begin{align}
\beta_{b}^{a}  &  =\digamma_{b}^{a}-\mathfrak{f}_{b}^{a}\\
\mathfrak{f}_{b}^{a}  &  =\frac{1}{2}\mathfrak{f}_{bcd}^{a}\theta^{c}%
\theta^{d}\nonumber
\end{align}
where $\mathfrak{f}_{b}^{a}\theta^{b}=0$ and therefore the functions
$\mathfrak{f}_{bcd}^{a}$ satisfy the conditions%
\begin{align}
\mathfrak{f}_{abcd}  &  =\mathfrak{f}_{[ab]cd}=\mathfrak{f}_{ab[cd]},\\
\mathfrak{f}_{[bcd]}^{a}  &  =0.\nonumber
\end{align}
Hence the first of the generalized Cartan equations, Eq.(11), reduces, in the
type $N=1$ case, to the statement that $\tau^{a}$ is the torsion two-form of
the $\mathfrak{so}(p,q)$connection given by $\alpha_{b}^{a}$ and the second
equation is just the covariant exterior derivative of the first and can be
re-expressed as $D\tau^{a}=\digamma_{b}^{a}\theta^{b}$. \ The only restriction
the type $N=1$ generalized Cartan equations place on the generalized
connection and curvature is that%
\begin{align}
\mathbf{A}_{b}^{a}  &  =\alpha_{b}^{a}-(\digamma_{b}^{a}-\mathfrak{f}_{b}%
^{a})\mathbf{m,}\\
\mathbf{F}_{b}^{a}  &  =\mathfrak{f}_{b}^{a}+D\mathfrak{f}_{b}^{a}%
\mathbf{m,}\nonumber
\end{align}
where the components of the two-form $\mathfrak{f}_{b}^{a}$ satisfy Eq.(33).

The type $N=1$ generalized connection is flat if and only if $\beta_{b}%
^{a}=\digamma_{b}^{a}$ and therefore $\mathfrak{f}_{b}^{a}=0$ and
\begin{equation}
\mathbf{A}_{b}^{a}=\alpha_{b}^{a}-\digamma_{b}^{a}\mathbf{m.}%
\end{equation}
\ As in the previous section, by expressing the generalized forms in terms of
their expansions in ordinary forms, it can be shown that such a flat
connection can be expressed as%
\begin{equation}
\mathbf{A}_{b}^{a}=(\mathbf{L}^{-1})_{c}^{a}d\mathbf{L}_{b}^{c}\mathbf{,}%
\end{equation}
where here the matrix ($\mathbf{L}_{b}^{a})\in SO(p,q)_{(1)}$ and%
\begin{equation}
\mathbf{L}_{b}^{a}\mathbf{=}\delta_{b}^{a}-\alpha_{b}^{a}\mathbf{m.}%
\end{equation}
It then again follows from the generalized Cartan equations, and the fact that
closed type $N\geqq1$ generalized forms are exact, that%
\begin{equation}
\mathbf{\mathbf{e}}^{a}=(\mathbf{L}^{-1})_{b}^{a}d\mathbf{x}^{b}%
\end{equation}
where%
\begin{equation}
\mathbf{x}^{a}=\xi^{a}+(d\xi^{a}-\theta^{a})\mathbf{m}\text{,}%
\end{equation}
and $\xi^{a}$ are arbitrary functions.

To summarize: the general solution of the type $N=1$ generalized Cartan
equations, Eqs.(30), with a flat generalized connection, Eq.(35), can be
written, modulo gauge transformations, in the form%
\begin{align}
\mathbf{\mathbf{e}}^{a}  &  =(\mathbf{L}^{-1})_{b}^{a}\mathbf{dx}^{b},\\
\mathbf{A}_{b}^{a}  &  =(\mathbf{L}^{-1})_{c}^{a}d\mathbf{L}_{b}^{c},\nonumber
\end{align}
where $\mathbf{L}_{b}^{a}$ is given by Eq.(37) and $\mathbf{x}^{a}$ is given
by Eq.(39).

Hence any ordinary co-frame $\{\theta^{a}\},$ for an ordinary metric
$\eta_{ab}\theta^{a}\otimes\theta^{b}$, with an $\mathfrak{so(p,q)}$-valued
connection $\alpha_{b}^{a}$ and torsion two-form $\tau^{a\text{ }}$can be
expressed in flat terms as in this equation. \ An alternative way of
demonstrating that in the case where $\alpha_{b}^{a}$ is the Levi-Civita
connection $\omega_{b}^{a}$ and the torsion is zero is the following. The
ordinary Cartan equations, Eq.(2) and (3), are satisfied if and only if the
following equations hold%
\begin{align}
d\theta^{a}+\mathbf{A}_{b}^{a}\theta^{b}  &  =0,\nonumber\\
\mathbf{A}_{b}^{a}  &  =\omega_{b}^{a}-\Omega_{b}^{a}\mathbf{m,}%
\end{align}
where the generalized connection $\mathbf{A}_{b}^{a}$ is flat. \ These
modified Cartan equations incorporate both the ordinary Cartan equations and
the first Bianchi identities and can be regarded as a special case of Eq.(11).
$\ $Since the curvature of the connection is zero it follows,as above, that
$\mathbf{A}_{b}^{a}=(\mathbf{L}^{-1})_{c}^{a}d\mathbf{L}_{b}^{c}$, where
$(\mathbf{L}_{b}^{a})\mathbf{=}\in SO(p,q)_{(1)}$ is given by%
\begin{equation}
\mathbf{L}_{b}^{a}=(\delta_{b}^{a}-\omega_{b}^{a}\mathbf{m}).
\end{equation}
Hence the ordinary co-frame can be expressed as\
\begin{equation}
\theta^{a}=(\mathbf{L}^{-1})_{b}^{a}\mathbf{dx}^{b},
\end{equation}
and the generalized form (gf-) coordinates $\{\mathbf{x}^{b}\}$ are the
generalized zero-forms
\begin{equation}
\mathbf{x}^{b}=\xi^{b}-(\theta^{b}-d\xi^{b})\mathbf{m},
\end{equation}
where $\xi^{b}$ are arbitrary functions, cf \cite{rob1}. \ Consequently every
ordinary metric can be formally represented by a line element $ds^{2}%
=\eta_{ab}d\mathbf{x}^{a}\otimes d\mathbf{x}^{b}$. \ These results are
co-frame covariant under an ordinary gauge transformation, like those in
Eq.(6), where $\theta^{a}\rightarrow\theta_{1}^{a}$ and $\omega_{b}%
^{a}\rightarrow\omega_{1b}^{a}$. \ That is%
\begin{equation}
\theta_{1}^{a}=(\mathbf{L}_{1}^{-1})_{b}^{a}d\mathbf{x}_{1}^{b}.
\end{equation}
where%
\begin{align}
(\mathbf{L}_{1}^{-1})_{b}^{a}  &  =\delta_{b}^{a}+\omega_{1b}^{a}%
\mathbf{m},\text{ }\mathbf{L}_{1b}^{a}=\delta_{b}^{a}-\omega_{1b}%
^{a}\mathbf{m},\\
\mathbf{x}_{1}^{a}  &  =(_{c}\mathbf{L}^{-1})_{b}^{a}\mathbf{x}^{b}.\nonumber
\end{align}
Here $(_{c}\mathbf{L})_{b}^{a}$ satisfies Eq.(14) and is closed (and therefore
exact). \ Written in the format of Eq.(15) $(_{c}\mathbf{L})_{b}^{a}$ and its
inverse are given by%
\begin{align}
(_{c}\mathbf{L})_{b}^{a}  &  =[\delta_{c}^{a}+(dL_{d}^{a})(L^{-1})_{c}%
^{d}\mathbf{m}](L)_{b}^{c},\\
(_{c}\mathbf{L}^{-1})_{b}^{a}  &  =[\delta_{c}^{a}-(L^{-1})_{d}^{a}dL_{c}%
^{d}\mathbf{m}](L^{-1})_{b}^{c}.\nonumber
\end{align}
\ The overall freedom in the choice of gf-coordinates $\mathbf{x}^{a}$ is
given by the affine group of generalized transformations
\begin{equation}
\mathbf{x}^{a}\rightarrow(_{c}\mathbf{L}^{-1})_{b}^{a}\mathbf{x}^{b}%
+_{c}\mathbf{p}^{a}%
\end{equation}
where the zero-forms $_{c}\mathbf{L}_{b}^{a}\in\mathbf{G}_{(1)}$ and the
generalized translations $_{c}\mathbf{p}^{a}$ are closed, $d_{c}\mathbf{L}%
_{b}^{a}=d_{c}\mathbf{p}^{a}=0$. \ The translation freedom can be used to set
$\xi^{a}=0$ in Eq.(44), a condition preserved by ordinary gauge transformations.

If $\{\theta^{a}\}$ is an ordinary co-frame satisfying the ordinary Cartan
equations as above so that it be expressed in terms of type $N=1$ generalized
forms by Eq.(43), that is $\theta^{a}=(\mathbf{L}^{-1})_{b}^{a}\mathbf{dx}%
^{b}$, any ordinary form can be expressed as an expansion in terms of the
exterior products of\ the co-frame of exact forms $\{\mathbf{dx}^{a}\}$.

\section{Lorentzian four-metrics, Cartan's structure equations and Einstein's
vacuum field equations}

In this section the formalism above will be adapted to the particular case of
Lorentzian four-metrics. The standard two component spinor formalism which
will be used will be reviewed and then extended, \cite{pen1}, \cite{rob9}.

Let $g$ be an ordinary four-metric of signature (1,3) on a real four
dimensional manifold $M$ \ with line element%
\begin{equation}
ds^{2}=\eta_{ab}\theta^{a}\otimes\theta^{b}.
\end{equation}
\ The structure group is the proper orthochronous Lorentz group $SO(1,3)$ and
is isomorphic to $SL(2,C)\times\overline{Sl(2,C)}/\mathbb{Z}_{2}$. \ Hence for
a real Lorentz transformation $L_{b}^{a}\in$ $SO(1,3)$%
\begin{equation}
L_{b}^{a}\longleftrightarrow L_{B}^{A}L_{B^{\prime}}^{A^{\prime}}%
\end{equation}
where $L_{B}^{A}$ and $L_{B^{\prime}}^{A^{\prime}}$ are complex conjugate
$SL(2,C)$ transformations,
\begin{equation}
L_{C}^{A}L_{D}^{B}\epsilon_{AB}=\epsilon_{CD}%
\end{equation}
where $\epsilon_{AB}$ is the skew symmetric spinor with $\epsilon_{01}=1,$ and
similarly for the complex conjugates. \ In terms of this formalism the line
element is%
\begin{equation}
ds^{2}=\epsilon_{AB}\epsilon_{A^{\prime}B^{\prime}}\theta^{AA^{\prime}}%
\otimes\theta^{BB^{\prime}}%
\end{equation}
where $\theta^{AA^{\prime}}$ denotes a Hermitian matrix-valued co-frame.

For such metrics the ordinary first Cartan structure equations can be
re-written as%
\begin{equation}
D\theta^{AA^{\prime}}\equiv d\theta^{AA^{\prime}}+\omega_{B}^{A}%
\theta^{BA^{\prime}}+\omega_{B^{\prime}}^{A^{\prime}}\theta^{AB^{\prime}}=0.
\end{equation}
The $\mathfrak{sl(2,C)}$-valued connection one-form $\omega_{B}^{A}$ and its
complex conjugate $\omega_{B^{\prime}}^{A^{\prime}}$ correspond, respectively,
to the anti self-dual and self-dual parts of the Levi-Civita connection one
form $\omega_{b}^{a}\longleftrightarrow\delta_{B^{\prime}}^{A^{\prime}}%
\omega_{B}^{A}+\delta_{B}^{A}\omega_{B^{\prime}}^{A^{\prime}}$.

The second Cartan equations are%
\begin{align}
\Omega_{B}^{A}  &  =d\omega_{B}^{A}+\omega_{C}^{A}\omega_{B}^{C}\\
\Omega_{B^{\prime}}^{A^{\prime}}  &  =d\omega_{B^{\prime}}^{A^{\prime}}%
+\omega_{C^{\prime}}^{A^{\prime}}\omega_{B^{\prime}}^{C^{\prime}},\nonumber
\end{align}
first for the anti self-dual part of the connection and curvature and second
for the self dual part, and $\Omega_{b}^{a}\longleftrightarrow\delta
_{B^{\prime}}^{A^{\prime}}\Omega_{B}^{A}+\delta_{B}^{A}\Omega_{B^{\prime}%
}^{A^{\prime}}$.

The first and second Bianchi identities are%
\begin{align}
\Omega_{B}^{A}\theta^{BA^{\prime}}+\Omega_{B^{\prime}}^{A^{\prime}}%
\theta^{AB^{\prime}}  &  =0,\\
D\Omega_{B}^{A}  &  =D\Omega_{B^{\prime}}^{A^{\prime}}=0,\nonumber
\end{align}
where $D$ denotes the covariant exterior derivative with respect to the
relevant anti self-dual or self-dual connection.

Under a gauge transformation by $L_{b}^{a}\longleftrightarrow L_{B}%
^{A}L_{B^{\prime}}^{A^{\prime}}$%
\begin{align}
\theta^{AA^{\prime}}  &  \rightarrow\theta_{1}^{AA^{\prime}}=(L^{-1}%
)_{B^{\prime}}^{A^{\prime}}(L^{-1})_{B}^{^{A}}\theta^{BB^{\prime}},\\
\omega_{B}^{A}  &  \rightarrow\omega_{1B}^{A}=(L^{-1})_{C}^{A}dL_{B}%
^{C}+(L^{-1})_{C}^{A}\text{ }\omega_{D}^{C}L_{B}^{D}\nonumber\\
\omega_{B^{\prime}}^{A^{\prime}}  &  \rightarrow\omega_{1B^{\prime}%
}^{A^{\prime}}=(L^{-1})_{C^{\prime}}^{A^{\prime}}dL_{B^{\prime}}^{C^{\prime}%
}+(L^{-1})_{C^{\prime}}^{A^{\prime}}\text{ }\omega_{D^{\prime}}^{C^{\prime}%
}L_{B^{\prime}}^{D^{\prime}}\nonumber\\
\Omega_{B}^{A}  &  \rightarrow\Omega_{1B}^{A}=(L^{-1})_{C}^{A}\Omega_{D}%
^{C}L_{B}^{D};\text{ }\Omega_{B^{\prime}}^{A^{\prime}}\rightarrow
\Omega_{1B^{\prime}}^{A^{\prime}}=(^{-}L^{-1})_{C^{\prime}}^{A^{\prime}}%
\Omega_{D^{\prime}}^{C^{\prime}}L_{B^{\prime}}^{D^{\prime}}.\nonumber
\end{align}
Einstein's vacuum field equations, with zero cosmological constant, are%
\begin{equation}
\Omega_{B}^{A}\theta^{BA^{\prime}}=0.
\end{equation}

The corresponding generalized form equations are obtained in the obvious way
and the Cartan equations are%
\begin{align}
\mathbf{De}^{AA^{\prime}}  &  \equiv d\mathbf{e}^{AA^{\prime}}+\mathbf{A}%
_{B}^{A}\mathbf{e}^{BA^{\prime}}+\mathbf{A}_{B^{\prime}}^{A^{\prime}%
}\mathbf{e}^{AB^{\prime}}=0,\\
\mathbf{F}_{B}^{A}  &  =d\mathbf{A}_{B}^{A}+\mathbf{A}_{C}^{A}\mathbf{A}%
_{B}^{C},\nonumber\\
\mathbf{F}_{B^{\prime}}^{A^{\prime}}  &  =d\mathbf{A}_{B^{\prime}}^{A^{\prime
}}+\mathbf{A}_{C^{\prime}}^{A^{\prime}}\mathbf{A}_{B^{\prime}}^{C^{\prime}%
},\nonumber
\end{align}
with first and second Bianchi identities%
\begin{align}
\mathbf{F}_{B}^{A}\mathbf{e}^{BA^{\prime}}+\mathbf{F}_{B^{\prime}}^{A^{\prime
}}\mathbf{e}^{AB^{\prime}}  &  =0,\\
\mathbf{DF}_{B}^{A}  &  =\mathbf{DF}_{B^{\prime}}^{A^{\prime}}=0,\nonumber
\end{align}
The generalized gauge transformations are as in Eq.(56) but with the
generalized forms replacing the ordinary forms. \ It can be verified by
straightforward calculation that if $\mathbf{L}_{b}^{a}$ is a generalized
Lorentz gauge transformation, satisfying $\mathbf{L}_{b}^{a}\eta
_{ac}\mathbf{L}_{d}^{c}=\eta_{bd},$ then, just as for ordinary groups,
$\mathbf{L}_{b}^{a}\longleftrightarrow\mathbf{L}_{B}^{A}\mathbf{L}_{B^{\prime
}}^{A^{\prime}}.$ \ Here $\mathbf{L}_{B}^{A}$ is a generalized $SL(2,C)$
transformations satisfying $\mathbf{L}_{C}^{A}\mathbf{L}_{D}^{B}\epsilon
_{AB}=\epsilon_{CD}$ and similarly for its complex conjugate $\mathbf{L}%
_{B^{\prime}}^{A^{\prime}}$. $\ \mathbf{L}_{b}^{a}$ is determined, as in
Eq.(15), by $L_{b}^{a},$ $_{1}\lambda_{b}^{a},$ $_{2}\lambda_{b}^{a},$
$\lambda_{b}^{a}$ where these are, respectively, a $SO(1,3)$-valued function,
two ordinary $\mathfrak{so(}1\mathfrak{,}3\mathfrak{)}$-valued one-forms and
an ordinary $\mathfrak{so(}1\mathfrak{,}3\mathfrak{)}$ -valued two-form. \ The
generalized $SL(2,C)$\ transformations $\{\mathbf{L}_{B}^{A}\}$ are determined
by functions and forms $L_{B}^{A},$ $_{1}\lambda_{B}^{A},$ $_{2}\lambda
_{B}^{A},$ $\lambda_{B}^{A}$,where these are respectively a $SL(2,C)$-valued
function, two ordinary $\mathfrak{sl(}2,C)$-valued one-forms and an ordinary
$\mathfrak{sl(}2,C\mathfrak{)}$ -valued two-form, and similarly for its
complex conjugate. \ Furthermore $L_{b}^{a}\leftrightarrow L_{B}%
^{A}L_{B^{\prime}}^{A^{\prime}},$ $_{1}\lambda_{b}^{a}\longleftrightarrow$
$\delta_{B^{\prime}}^{A^{\prime}}$ $_{1}\lambda_{B}^{A}+\delta_{B}^{A}$
$_{1}\lambda_{B^{\prime}}^{A^{\prime}}$, $_{2}\lambda_{b}^{a}%
\longleftrightarrow$ $\delta_{B^{\prime}}^{A^{\prime}}$ $_{2}\lambda_{B}%
^{A}+\delta_{B}^{A}$ $_{2}\lambda_{B^{\prime}}^{A^{\prime}}$ and $\lambda
_{b}^{a}\longleftrightarrow\delta_{B^{\prime}}^{A^{\prime}}$ $\lambda_{B}%
^{A}+\delta_{B}^{A}$ $\lambda_{B^{\prime}}^{A^{\prime}}$.

\section{Generalized forms and Einstein's vacuum field equations for
space-time metrics}

Einstein's vacuum field equations, Eqs.(57), can be combined with
Cartan's\ first structure equations, Eq.(53), into the generalized forms
equation%
\begin{equation}
\mathbf{D}\theta^{AA^{\prime}}\equiv d\theta^{AA^{\prime}}+\mathbf{A}_{B}%
^{A}\theta^{BA^{\prime}}+\mathbf{A}_{B^{\prime}}^{A^{\prime}}\theta
^{AB^{\prime}}=0.
\end{equation}
Here $\mathbf{A}_{B}^{A}$ and $\mathbf{A}_{B^{\prime}}^{A^{\prime}}$
respectively denote complex conjugate generalized connection one-forms, taking
values in the Lie algebra of the generalized $SL(2,C)$ as above. \ They are
given by%
\begin{equation}
\mathbf{A}_{B}^{A}=\text{ }\omega_{B}^{A}-\Omega_{B}^{A}\mathbf{m}\text{;
}\mathbf{A}_{B^{\prime}}^{A^{\prime}}=\omega_{B^{\prime}}^{A^{\prime}}%
-\Omega_{B^{\prime}}^{A^{\prime}}\overline{\mathbf{m}}%
\end{equation}
and have zero generalized curvature, that is they are flat.

Hence Einstein's vacuum field equations are satisfied if and only if Eqs.(60)
and (61) hold.

Consequently when type $N=2$ one-forms $\{\mathbf{e}^{AA^{\prime}}\}$, satisfy
the generalized first Cartan equations,%
\begin{equation}
\mathbf{De}^{AA^{\prime}}\equiv d\mathbf{e}^{AA^{\prime}}+\mathbf{A}_{B}%
^{A}\mathbf{e}^{BA^{\prime}}+\mathbf{A}_{B^{\prime}}^{A^{\prime}}%
\mathbf{e}^{AB^{\prime}}=0,
\end{equation}
with $\mathbf{A}_{B}^{A}$ and $\mathbf{A}_{B^{\prime}}^{A^{\prime}}$ flat
generalized $\mathfrak{sl(2,C)}$-valued connections%
\begin{align}
\mathbf{A}_{B}^{A}  &  =\alpha_{B}^{A}-\digamma_{B}^{A}\mathbf{m,}\\
\digamma_{B}^{A}  &  =d\alpha_{B}^{A}+\alpha_{C}^{A}\alpha_{B}^{C}\nonumber
\end{align}
(and similarly for the complex conjugates), and $\{\mathbf{e}^{AA^{\prime}}\}$
constitutes an ordinary co-frame, that is%
\begin{equation}
\mathbf{e}^{AA^{\prime}}=\theta^{AA^{\prime}}%
\end{equation}
then $\{\theta^{AA^{\prime}}\}$ determines a solution of Einstein's vacuum
field equations. \ Moreover $\alpha_{B}^{A}=\omega_{B}^{A}.$

Because the connections in Eq.(61) are flat they can be expressed in terms of
elements of the (pointwise) groups of generalized forms$\{\mathbf{L}_{B}%
^{A}\}$and $\{\mathbf{L}_{B^{\prime}}^{A^{\prime}}\}$%

\begin{equation}
\mathbf{A}_{B}^{A}=(\mathbf{L}^{-1})_{C}^{A}d\mathbf{L}_{B}^{C};\text{
}\mathbf{A}_{B^{\prime}}^{A^{\prime}}=(\mathbf{L}^{-1})_{C^{\prime}%
}^{A^{\prime}}d\mathbf{L}_{B^{\prime}}^{C^{\prime}}%
\end{equation}

where
\begin{align}
\mathbf{L}_{B}^{A}  &  =\delta_{B}^{A}-\omega_{B}^{A}\mathbf{m}\text{;
}(\mathbf{L}^{-1}\mathbf{)}_{B}^{A}=\delta_{B}^{A}+\omega_{B}^{A}\mathbf{m,}\\
\mathbf{L}_{B^{\prime}}^{A^{\prime}}  &  =\delta_{B^{\prime}}^{A^{\prime}%
}-\omega_{B^{\prime}}^{A^{\prime}}\overline{\mathbf{m}}\text{; }%
(\mathbf{L}^{-1}\mathbf{)}_{B^{\prime}}^{A^{\prime}}=\delta_{B^{\prime}%
}^{A^{\prime}}+\omega_{B^{\prime}}^{A^{\prime}}\overline{\mathbf{m}}.\nonumber
\end{align}
\ Hence a co-frame $\theta^{a}$ determines solutions of the vacuum field
equations if and only if, modulo ordinary gauge transformations, it can be
expressed as
\begin{equation}
\theta^{AA^{\prime}}=(\mathbf{L}^{-1})_{b}^{a}d\mathbf{x}^{b}=(\mathbf{L}%
^{-1})_{B^{\prime}}^{A^{\prime}}(\mathbf{L}^{-1})_{B}^{A}d\mathbf{x}%
^{BB^{\prime}},
\end{equation}
where $\mathbf{L}_{b}^{a}\longleftrightarrow\mathbf{L}_{B}^{A}\mathbf{L}%
_{B^{\prime}}^{A^{\prime}}$ and the generalized zero-forms\ $\mathbf{x}%
^{AA^{\prime}}$ are (non-unique) gf-coordinates. \ The corresponding line
element, $ds^{2}=\eta_{ab}\theta^{a}\otimes\theta^{b}$ can be formally
expressed as%
\begin{equation}
ds^{2}=\eta_{ab}d\mathbf{x}^{a}\otimes d\mathbf{x}^{b}=\epsilon_{AB}%
\epsilon_{A^{\prime}B^{\prime}}d\mathbf{x}^{AA^{\prime}}\otimes d\mathbf{x}%
^{BB^{\prime}}.
\end{equation}

These results are co-frame covariant in the following sense. \ If under an
ordinary gauge transformation%
\begin{equation}
\theta^{AA^{\prime}}\rightarrow\theta_{1}^{AA^{\prime}}=(L^{-1})_{B^{\prime}%
}^{A^{\prime}}(L^{-1})_{B}^{A}\theta^{BB^{\prime}},
\end{equation}
and so on as in Eq.(56), it is a straightforward matter to show that%
\begin{equation}
\theta_{1}^{AA^{\prime}}=(\mathbf{L}_{1}^{-1})_{B^{\prime}}^{A^{\prime}%
}(\mathbf{L}_{1}^{-1})_{B}^{A}d\mathbf{x}_{1}^{BB^{\prime}}=(\mathbf{L}%
_{1}^{-1})_{b}^{a}d\mathbf{x}_{1}^{b}.
\end{equation}
where%
\begin{align}
\mathbf{L}_{1B}^{A}  &  =\delta_{B}^{A}-\text{ }\omega_{1B}^{A}\mathbf{m,}%
\text{ }(\mathbf{L}_{1}^{-1})_{B}^{A}=\delta_{B}^{A}+\text{ }\omega_{1B}%
^{A}\mathbf{m,}\\
\mathbf{L}_{1B^{\prime}}^{A^{\prime}}  &  =\delta_{B^{\prime}}^{A^{\prime}%
}-\text{ }\omega_{1B^{\prime}}^{A^{\prime}}\overline{\mathbf{m}}%
\mathbf{,}\text{ }(\mathbf{L}_{1}^{-1})_{B^{\prime}}^{A^{\prime}}%
=\delta_{B^{\prime}}^{A^{\prime}}+\text{ }\omega_{1B^{\prime}}^{A^{\prime}%
}\overline{\mathbf{m}}\mathbf{.}\nonumber
\end{align}

The transformed gf-coordinates are given by
\begin{equation}
\mathbf{x}_{1}^{a}=(_{c}\mathbf{L}^{-1})_{b}^{a}\mathbf{x}^{b}%
\longleftrightarrow(_{c}\mathbf{L}^{-1})_{B^{\prime}}^{A^{\prime}}%
(_{c}\mathbf{L}^{-1})_{B}^{A}\mathbf{x}^{BB^{\prime}}.
\end{equation}
Here $(_{c}\mathbf{L})_{b}^{a}$, $_{c}\mathbf{L}_{B}^{A}$ \ and $_{c}%
\mathbf{L}_{B^{\prime}}^{A^{\prime}}$ are closed (and therefore exact) and
satisfy Eq.(41) and the generalized forms versions of Eq.(51) and its complex
conjugate. \ Written in the format of Eq.(15) they are%
\begin{equation}
_{c}\mathbf{L}_{B}^{A}=[\delta_{C}^{A}+(dL_{D}^{A})(L^{-1})_{C}^{D}%
\mathbf{m}]L_{B}^{C},
\end{equation}
\ with inverse%
\begin{equation}
(_{c}\mathbf{L}^{-1})_{B}^{A}=[\delta_{c}^{a}-(L^{-1})_{D}^{A}dL_{C}%
^{D}\mathbf{m}](L^{-1})_{B}^{C},
\end{equation}
and similarly for $(_{c}\mathbf{L})_{B^{\prime}}^{A^{\prime}}$. \ There is
also a generalized translation freedom%
\begin{equation}
\mathbf{x}^{AA^{\prime}}\rightarrow\mathbf{x}^{AA^{\prime}}+d\mathbf{b}%
^{AA^{\prime}}%
\end{equation}
arising from the fact that $d\mathbf{x}^{AA^{\prime}}=d(\mathbf{x}%
^{AA^{\prime}}+d\mathbf{b}^{AA^{\prime}})$ where $\mathbf{b}^{AA^{\prime}}$ is
a generalized form of degree minus one.

Hence the overall freedom in the choice of gf coordinate is given by a
generalized Poincar\'{e} transformation%
\begin{equation}
\mathbf{x}^{AA^{\prime}}\rightarrow(_{c}\mathbf{L}^{-1})_{B}^{A}%
(_{c}\mathbf{L}^{-1})_{B^{\prime}}^{A^{\prime}}\mathbf{x}^{BB^{\prime}}+\text{
}_{c}\mathbf{p}^{AA^{\prime}},
\end{equation}
where the generalized Lorentz transformation $_{c}\mathbf{L}_{b}%
^{a}\leftrightarrow$ $_{c}\mathbf{L}_{B}^{A}{}_{c}\mathbf{L}_{B^{\prime}%
}^{A^{\prime}}$ and the generalized translation $_{c}\mathbf{p}^{a}%
\leftrightarrow_{c}\mathbf{p}^{AA^{\prime}}$ satisfy $d$ $_{c}\mathbf{L}%
_{b}^{a}=d_{c}\mathbf{p}^{a}=0.$

To summarize: an ordinary co-frame, $\theta^{AA^{\prime}},$ and connections
$\omega_{B}^{A}$ and $\omega_{B^{\prime}}^{A^{\prime}}$, determine a
Lorentzian four-metric and Levi-Civita connection, $\omega_{b}^{a}%
\longleftrightarrow\delta_{B^{\prime}}^{A^{\prime}}\omega_{B}^{A}+\delta
_{B}^{A}\omega_{B^{\prime}}^{A^{\prime}}$, satisfying Einstein's empty space
equations if and only if the co-frame can be expressed in terms of generalized
forms as in Eqs.(66) and (67).

In order to express this result in terms of ordinary forms, and determine a
choice of $\mathbf{x}^{AA^{\prime}}$, let the expansion of the generalized
zero-form coordinates $\mathbf{x}^{AA^{\prime}}$ be%
\begin{equation}
\mathbf{x}^{AA^{\prime}}=\xi^{AA^{\prime}}-\mu^{AA^{\prime}}\mathbf{m}%
-\overline{\mu}^{AA^{\prime}}\overline{\mathbf{m}}+i\nu^{AA^{\prime}%
}\mathbf{m}\overline{\mathbf{m}},
\end{equation}
where $\xi^{AA^{\prime}}$, $\mu^{AA^{\prime}}$ and $\nu^{AA^{\prime}}$
respectively correspond spinorially to real ordinary zero-forms, complex
ordinary one-forms and ordinary real two-forms. \ This expansion can be
changed by using the translational freedom of Eq.(75). \ By using this freedom
the expansion in Eq.(77) can be simplified to%
\begin{equation}
\mathbf{x}^{AA^{\prime}}=-\mu^{AA^{\prime}}\mathbf{m}-\mu^{AA^{\prime}%
}\overline{\mathbf{m}}+i\nu^{AA^{\prime}}\mathbf{m}\overline{\mathbf{m}},
\end{equation}
where now the one-forms $\mu^{a}\leftrightarrow\mu^{AA^{\prime}}$ are real,
that is $(\mu^{AA^{\prime}})$ is a hermitian matrix-valued one-form. \ This
simplification is preserved under the ordinary gauge transformations of
Eq.(56). \ Under such gauge transformations
\begin{equation}
\mathbf{x}^{AA^{\prime}}\rightarrow\mathbf{x}_{1}^{AA^{\prime}}=-\mu
_{1}^{AA^{\prime}}\mathbf{m}-\mu_{1}^{AA^{\prime}}\overline{\mathbf{m}}%
+i\nu_{1}^{AA^{\prime}}\mathbf{m}\overline{\mathbf{m}}%
\end{equation}
where%
\begin{align}
\mu_{1}^{AA^{\prime}}  &  =(L^{-1})_{B^{\prime}}^{A^{\prime}}(L^{-1})_{B}%
^{A}\mu^{BB^{\prime}}\\
\nu_{1}^{AA^{\prime}}  &  =(L^{-1})_{B^{\prime}}^{A^{\prime}}(L^{-1})_{B}%
^{A}\nu^{BB^{\prime}}+i(L^{-1})_{B^{\prime}}^{A^{\prime}}d[(L^{-1})_{B}%
^{A}]\mu^{BB^{\prime}}\nonumber\\
&  -i[d(L^{-1})_{B^{\prime}}^{A^{\prime}}](L^{-1})_{B}^{A}\mu^{BB^{\prime}%
}.\nonumber
\end{align}

The expanded form of $\mathbf{e}^{AA^{\prime}}=(\mathbf{L}^{-1})_{B^{\prime}%
}^{A^{\prime}}(\mathbf{L}^{-1})_{B}^{A}d\mathbf{x}^{BB^{\prime}}$ is, using
Eq.(66) and (78), now%
\begin{align}
\mathbf{e}^{AA^{\prime}}  &  =2\mu^{AA^{\prime}}\\
&  -[d\mu^{AA^{\prime}}+2\omega_{B}^{A}\mu^{BA^{\prime}}+i\nu^{AA^{\prime}%
}]\mathbf{m}\nonumber\\
&  -[d\mu^{AA^{\prime}}+2\omega_{B^{\prime}}^{A^{\prime}}\mu^{AB^{\prime}%
}-i\nu^{AA^{\prime}}]\overline{\mathbf{m}}\nonumber\\
&  +[i(d\nu^{AA^{\prime}}+\omega_{B^{\prime}}^{A^{\prime}}\nu^{AB^{\prime}%
}+\omega_{B}^{A}\nu^{BA^{\prime}})+\omega_{B^{\prime}}^{A^{\prime}}%
d\mu^{AB^{\prime}}-\omega_{B}^{A}d\mu^{BA^{\prime}}-2\omega_{B}^{A}%
\omega_{B^{\prime}}^{A^{\prime}}\mu^{BB^{\prime}}]\mathbf{m}\overline
{\mathbf{m}}.\nonumber
\end{align}
It follows then that $\mathbf{e}^{AA^{\prime}}$ are ordinary forms if and only
if%
\begin{equation}
\mathbf{e}^{AA^{\prime}}=2\mu^{AA^{\prime}}%
\end{equation}
and
\begin{align}
D\mu^{AA^{\prime}}  &  =0,\\
\nu^{AA^{\prime}}  &  =i(\omega_{B}^{A}\mu^{BA^{\prime}}-\omega_{B^{\prime}%
}^{A^{\prime}}\mu^{AB^{\prime}})\nonumber\\
\Omega_{B}^{A}\mu^{BA^{\prime}}-\Omega_{B^{\prime}}^{A^{\prime}}%
\mu^{AB^{\prime}}  &  =0,\nonumber
\end{align}
and $D$ denotes the ordinary exterior covariant derivative with respect to
$\omega_{b}^{a}\longleftrightarrow(\delta_{B^{\prime}}^{A^{\prime}}\omega
_{B}^{A}+\delta_{B}^{A}\omega_{B^{\prime}}^{A^{\prime}})$. \ In terms of these
ordinary forms it follows from Eqs.(82) and (83) that, writing $\mu
^{AA^{\prime}}=\frac{1}{2}\theta^{AA^{\prime}}$,%
\begin{align}
D\theta^{AA^{\prime}}  &  =0,\\
\Omega_{B}^{A}\theta^{BA^{\prime}}-\Omega_{B^{\prime}}^{A^{\prime}}%
\theta^{AB^{\prime}}  &  =0,\nonumber
\end{align}
and hence the first Cartan equation and therefore the first Bianchi identity,
and Einstein's vacuum field equations, Eq.(57), are satisfied by\footnote{It
should be noted that this representation is different from the $N=1$
representation of \textit{any} metric as a flat gf-metric.}
\begin{equation}
ds^{2}=\epsilon_{AB}\epsilon_{A^{\prime}B^{\prime}}\mathbf{e}^{AA^{\prime}%
}\otimes\mathbf{e}^{BB^{\prime}}=\epsilon_{AB}\epsilon_{A^{\prime}B^{\prime}%
}\theta^{AA^{\prime}}\otimes\theta^{BB^{\prime}}=\epsilon_{AB}\epsilon
_{A^{\prime}B^{\prime}}\mathbf{dx}^{AA^{\prime}}\otimes\mathbf{dx}%
^{BB^{\prime}}.
\end{equation}

Conversely, if Einstein's vacuum field equations are satisfied by
$ds^{2}=\epsilon_{AB}\epsilon_{A^{\prime}B^{\prime}}\theta^{AA^{\prime}%
}\otimes\theta^{BB^{\prime}}$ so are Eqs.(82) \& (83) with
\begin{equation}
\mu^{AA^{\prime}}=\frac{1}{2}\theta^{AA^{\prime}};\text{ }\nu^{AA^{\prime}%
}=\frac{i}{2}(\omega_{B}^{A}\theta^{BA^{\prime}}-\omega_{B^{\prime}%
}^{A^{\prime}}\theta^{AB^{\prime}}),
\end{equation}
and a co-frame for the Ricci flat metric is given by%
\begin{equation}
\theta^{AA^{\prime}}=(\mathbf{L}^{-1}\mathbf{)}_{B^{\prime}}^{A^{\prime}%
}(\mathbf{L}^{-1}\mathbf{)}_{B}^{A}\mathbf{dx}^{BB^{\prime}\text{ }}%
=(\delta_{B^{\prime}}^{A^{\prime}}+\omega_{B^{\prime}}^{A^{\prime}}%
\overline{\mathbf{m}}\mathbf{)}(\delta_{B}^{A}+\omega_{B}^{A}\mathbf{m)dx}%
^{BB^{\prime}\text{ }}%
\end{equation}
where
\begin{equation}
\mathbf{x}^{AA^{\prime}\text{ }}=-\frac{1}{2}\theta^{AA^{\prime}}%
\mathbf{m}-\frac{1}{2}\theta^{AA^{\prime}}\overline{\mathbf{m}}+\frac{1}%
{2}[\omega_{B^{\prime}}^{A^{\prime}}\theta^{AB^{\prime}}-\omega_{B}^{A}%
\theta^{BA^{\prime}}]\mathbf{m}\overline{\mathbf{m}}.
\end{equation}
The Ricci flat metric can be formally expressed in terms of these
gf-coordinates as in Eq.(85).

\section{Discussion}

Previous work on generalized forms, Cartan's structure equations and ordinary
metric geometries has been extended. It has been shown that any ordinary
metric can be formally expressed as a flat type $N=1$ metric. \ Lorentzian
four-metrics were shown to be Ricci flat if and only if they can be expressed
as flat type $N=2$ metrics, but only by those belonging to a particular
sub-class. \ This was done by using degree minus one-forms to split the
self-dual and anti-self dual components of the ordinary Levi-Civita connection.

The approach taken in this paper can, with appropriate modifications, be
applied to other metric geometries, for example to holomorphic half-flat
metrics, four-metrics of non-Lorentzian signatures and geometries in other
dimensions. \ Dual formulations using generalized vector fields can also be
constructed. \ It would be interesting to explore the extent to which
gf-coordinates, of the type introduced in this paper, can be usefully employed
in the study of geometries and their physical applications.

\end{document}